\documentclass[12pt,preprint]{aastex}
\usepackage{graphicx}
\begin{document}
\def\la{\mathrel{\hbox{\rlap{\hbox{\lower4pt\hbox{$\sim$}}}\hbox{$<$}}}}
\def\ga{\mathrel{\hbox{\rlap{\hbox{\lower4pt\hbox{$\sim$}}}\hbox{$>$}}}}
\def\lam{$\lambda$}
\def\kms{km~s$^{-1}$}
\def\vphot{$v_{phot}$}
\def\ang{~\AA}

\title {Reading the Spectra of the Most Peculiar Type~Ia Supernova 2002cx}

\author {David Branch\altaffilmark{1}, E.~Baron\altaffilmark{1},
   R.~C.~Thomas\altaffilmark{2}, D.~Kasen\altaffilmark{2},
   Weidong~Li\altaffilmark {3}, and
   Alexei~V.~Filippenko\altaffilmark{3}}

\altaffiltext{1}{Department of Physics and Astronomy, University of
Oklahoma, Norman, Oklahoma 73019, USA; branch@nhn.ou.edu}

\altaffiltext{2}{Lawrence Berkeley National Laboratory, 1~Cyclotron
  Road, Berkeley, CA 94720--8158}

\altaffiltext{3}{Department of Astronomy, 601~Campbell Hall,
University of California, Berkeley, CA 94720--3411}

\begin{abstract}

In spite of the apparent lack of Si~II and S~II features in its
spectra, SN~2002cx was classified as a peculiar Type~Ia supernova
(SN~Ia) on the basis of its overall photometric and spectroscopic
behavior.  Spectra obtained near maximum light contained Fe~III
features, as in SN~1991T--like events, but the blueshifts of the
Fe~III absorptions were exceptionally low.  The luminosity also was
low.  We use the supernova synthetic--spectrum code {\bf Synow} to
study line identifications in SN~2002cx.  We find that the
maximum--light spectra appear to contain weak features of Si~II, S~II,
Si~III, and Ca~II, which strengthens the connection with
SN~1991T--like events.  We show that later spectra, obtained 12, 25,
and 56 days after maximum, consist of P--Cygni resonance--scattering
features due to permitted Fe~II and Co~II lines.  SN~2002cx had been
thought to have made the transition from a permitted--line to a
forbidden--line spectrum between 25 and 56 days.  Owing to the low
expansion velocities the postmaximum spectral features are narrower
and easier to identify than they are in other SNe~Ia.  SN~2002cx will
lead to improved line identifications in other SNe~Ia and clarify when
the transition from a permitted--line to a forbidden--line spectrum
occurs.  In the context of current SN~Ia explosion models, we suggest
that the properties of SN~2002cx may be consistent with 3D
deflagration models, which are not favored for normal SNe~Ia.

\end{abstract}

\keywords{supernovae: general -- supernovae: individual (SN~2002cx)}

\section{INTRODUCTION}

Supernova 2002cx was a very peculiar event.  Li et~al. (2003;
 hereafter L03) presented optical photometric and spectroscopic
 observations from which they concluded that
 SN~2002cx\footnote{SN~2002cx should not be confused with SN~2000cx, a
 peculiar SN~Ia of a different kind, also observed by Li et~al. (2001)
 and subjected to a synthetic--spectrum analysis by Branch
 et~al. (2004).} was the most peculiar known Type~Ia supernova
 (SN~Ia).  In spectra obtained near the time of maximum light the only
 line identifications were two conspicuous features due to Fe~III
 \lam4404 and \lam 5129, such as appear in peculiar SN~1991T--like
 events (Filippenko et~al. 1992a; see also Filippenko 1997 for a
 review of supernova spectra), but in SN~2002cx the Fe~III absorptions
 were blueshifted only by about 7000 \kms.  At maximum light SN~2002cx
 had a normal $B-V$ color, yet it was almost as subluminous as
 peculiar SN~1991bg--like events (Filippenko et~al. 1992b), which have
 much redder $B-V$ colors and completely different spectra.  L03 did
 not identify the characteristic SN~Ia features of Si~II and S~II at
 any phase.  In postmaximum spectra Fe~II features developed unusually
 early, and at unusually low velocities.  A 56--day postmaximum
 spectrum appeared to indicate that SN~2002cx had completed an early
 transition to the nebular phase.\footnote{In this paper we use a
 spectroscopic definition of nebular phase: the phase during which the
 spectrum is dominated by forbidden emission lines rather than by
 P--Cygni features of permitted lines.}  L03 concluded that no
 existing explosion model can account for the properties of SN~2002cx.

Because many of the spectral features were not identified by L03, and
because the low expansion velocities make the spectral features
relatively narrow and potentially identifiable, we have used the
parameterized synthetic--spectrum code {\bf Synow} to study line
identifications in selected spectra of SN~2002cx.

\section{ANALYSIS}

L03 presented eight spectra of SN~2002cx, ranging from 4~days before
to 56 days after the time of maximum light in the $B$ band.  The four
spectra selected for our study, obtained one day before and 12, 25,
and 56 days after maximum, are shown in Figure~1.  (The spectrum
labelled day~25 actually includes a small portion of a day~20
spectrum, from 9100 to 10,000\ang; the day~25 and day~20 spectra are
very similar in their extensive region of overlap.)  In all figures of
this paper the logarithm of the wavelength is plotted, to allow a fair
comparison of the widths of the spectral features at different
wavelengths.  Figure~1 shows that the day~$-1$ and day~12 spectra are
quite different, while the day~12, day~25, and day~56 spectra appear
similar in many respects.

In recent papers (Branch et~al. 2003, 2004) we have used {\bf Synow}
to interpret spectra of the normal Type~Ia SN~1998aq and the peculiar
Type~Ia SN~2000cx.  For discussions of the code, its input parameters,
and the way it is used, we refer the reader to these papers.

\subsection{The Day~$-1$ Spectrum}

The two earliest specta obtained by L03, 4~days and 1~day before
maximum, as well as the day $-4$ spectrum of the SN~1991T--like
SN~1997br (Li et~al. 1999) are compared in Figure~2.  The SN~1997br
spectrum has been redshifted by 3000 \kms\ to align the Fe~III
absorptions.  In addition to the two Fe~III features, many weaker
features appear in both of the SN~2002cx spectra (especially in the
blue) and therefore appear to be real.  We chose to study the day~$-1$
spectrum because most of its features are stronger than in the
day~$-4$ spectrum.

Figure~3 shows a comparison of the day~$-1$ spectrum with a synthetic
spectrum that has a velocity at the photosphere of \vphot\ = 7000
\kms, a blackbody continuum temperature of $T_{bb}=14,000$~K, and
contains lines of six ions: Fe~III, Si~II, Si~III, S~II, Ca~II, and
Ti~II.  The ion--specific input parameters --- the wavelength
($\lambda_{ref}$) and maximum optical depth ($\tau_{ref}$) of the
reference line; the minimum ($v_{min}$), maximum ($v_{max}$), and
optical--depth e--folding ($v_e$) velocities; and the excitation
temperature ($T_{exc}$) --- are listed in Table~1.  Following Branch
et~al. (2003), for each ion $T_{exc}$ is assigned the value at which
the reference--line optical depth reaches its maximum in
local--thermodynamic--equilibrium (LTE) calculations (Hatano
et~al. 1999), or 5000~K, whichever is higher, because some
reference--line optical depths peak only at very low temperatures.
(One exception to this rule for choosing $T_{exc}$ will be mentioned
in \S2.4.) Figure~3 shows that Si~II \lam6355, S~II \lam5468 and
\lam5654, Ca~II H\&K \lam3945, and Si~III \lam4550, all forming above
the same photospheric velocity as the Fe~III lines, can account
reasonably well for weak observed features.  (Ti~II \lam3760 should be
regarded as only one possible identification for the absorption near
3680\ang.)

Several of the observed features are not accounted for by the
synthetic spectrum of Figure~3. Some may be produced by weak Fe~III,
Si~III, and S~II lines: when the reference--line optical depths of
these ions are increased, thus making the strongest synthetic features
too strong, weaker features appear in the synthetic spectrum near some
of the observed ones.  Various effects, including partial covering of
the photosphere by these ions, and departures from LTE excitation of
their level populations, could cause the relative strengths of
observed weak and strong features to be greater than they are in a
{\bf Synow} spectrum.  Another possibility is that some of the
unidentified features are due to detached high--velocity
features\footnote{Detached refers to non--zero line optical depths
only above a velocity that exceeds the velocity at the photosphere.
Recently it has been realized that detached high--velocity features
are not uncommon in SN~Ia spectra (e.g., Branch 2004; Gerardy et
al. 2004).}: e.g., when we introduce line formation detached at
13,000~\kms, Fe~II lines and Ca~II H\&K can account for a few of the
observed absorptions.  However, we have not found a convincing way to
identify all of the observed features.  This raises the spectre of
line formation in multiple shells or clumps at various line--of--sight
velocities (Wang et~al. 2003; Kasen et~al. 2003; Thomas et~al 2004).

L03 showed that the early spectra of SN~2002cx were in some respects
like those of SN~1991T--like events, although with exceptionally low
absorption blueshifts.  Figure~4 shows a comparison of the day~$-4$
spectrum of SN~1997br (Li et~al. 1999) with a synthetic spectrum that
is exactly like the one that is compared to SN~2002cx in Figure~3,
except that \vphot\ has been increased from 7000 to 12,000 \kms.
Although there are discrepancies in Figure~4, the agreement is about
as good as it is in Figure~3, i.e., the ``sped up'' synthetic spectrum
for SN~2002cx resembles the spectrum of SN~1997br.  The near--maximum
spectra of SN~1991T--like events are known to contain weak Si~II,
S~II, and Ca~II features because these features were seen to
strengthen gradually in postmaximum spectra (Mazzali et~al. 1995; Li
et~al. 1999; Hatano et~al. 2002).  Together Figures~3 and 4 strongly
suggest the presence of weak Si~II, S~II, and Ca~II features in
SN~2002cx, at roughly the same optical depths as in SN~1997br.

\subsection{The Day~12 Spectrum}

L03 found that in the day~12 spectrum the Fe~III lines had vanished
and strong Fe~II lines had developed.  Figure~5 shows a comparison of
the day~12 spectrum with a synthetic spectrum that has \vphot\ = 7000
\kms, $T_{bb} = 9000$~K, and contains lines of five ions: Ca~II, Na~I,
Fe~II, Co~II, and Cr~II.  In the synthetic spectrum Na~I \lam 5892 and
Ca~II \lam3945 each produce just one feature.  Lines of Fe~II produce
most of the other observed features, but line blanketing by Co~II
lines is needed to get a reasonable fit in the blue.  Lines of Cr~II
also improve the fit in the blue.  Apart from the Ca~II feature, at
wavelengths shorter than 5500\ang\ the synthetic spectrum is a complex
blend of Fe~II, Co~II, and Cr~II lines.  The ion--specific input
parameters are in Table~2.  In the synthetic spectrum the Fe~II,
Co~II, and Cr~II lines are forming from 7000 \kms\ to an imposed
maximum velocity of 9000 \kms, with line optical depths that are
independent of velocity within this range ($v_e=\infty$); thus in the
synthetic spectrum the optical depths of the Fe~II, Co~II, and Cr~II
lines decrease discontinuously (and artificially) to zero at 9000
\kms. The Na~I and Ca~II lines have flat optical--depth distributions
from 7000 to 9000 \kms, but in addition (see Table~2) they have
exponentially decreasing ($v_e=2000$) optical depth above 9000 \kms.

L03 showed that in some respects the day~12 spectrum of SN~2002cx
resembled the spectra of other SNe~Ia at later epochs, about three
weeks postmaximum, but with lower absorption blueshifts in SN~2002cx.
We find that a synthetic spectrum like that of Figure~5 best resembles
three--week postmaximum spectra of other SNe~Ia when the velocity at
the photosphere and the maximum velocities are scaled up by a factor
of 9/7.

\subsection{The Day 25 Spectrum}

L03 showed that spectra obtained between 20 and 27 days in some
respects resembled other SNe~Ia at about six weeks postmaximum, again
with lower blueshifts in SN~2002cx.  Mysterious flux peaks between
6600 and 8000\ang\ appeared to be genuine emission lines rather than
P--Cygni features, perhaps signaling the beginning of an early
transition to the nebular phase (but see below).

Figure~6 shows a comparison of the day~25 spectrum with a synthetic
spectrum that has \vphot\ = 5000 \kms, $T_{bb}=5500$~K, and contains
lines of the same five ions used for the day~12 spectrum.  In this
synthetic spectrum Fe~II again produces many of the features.
However, Co~II lines not only improve the fit in the blue but they
also account for several individual features in the red.  This is the
first clear evidence for permitted Co~II features in the red part of
an SN~Ia spectrum.  Now that the observed spectrum extends to
sufficiently long wavelengths, the Ca~II triplet also is evident.  The
ion--specific parameters are in Table~3.  The Fe~II, Co~II, and Cr~II
features are forming between 5000 \kms\ and an imposed maximum
velocity of 7000 \kms\ while Na~I and Ca~II are forming from 5000 to
9000 \kms, all with flat optical--depth distributions.  The degree to
which Fe~II and Co~II can account for nearly all of the features in
the day~25 spectrum, including the apparent emission lines between
6600 and 8000\ang, indicates that (despite discrepancies such as the
insufficiently deep synthetic absorption near 6900\ang) our present
interpretation in terms of permitted--line P--Cygni features is
correct; there is no need to invoke the development of nebular--phase
emission lines.  Note also that the absorption near 7600\ang, commonly
attributed to O~I \lam7773 in early SN~Ia spectra, as well as in this
spectrum of SN~2002cx by L03, is accounted for by a blend of Fe~II
\lam7690, \lam7705, and \lam7732; this shows that when Fe~II lines are
strong even in the red, O~I \lam7773 is not necessarily required.  We
find that a synthetic spectrum like that of Figure~6 best resembles
six--week postmaximum spectra of other SNe~Ia when the input
velocities are scaled up by a factor of 9/5.

\subsection{The Day 56 Spectrum}

L03 considered the day~56 spectrum to be a nebular--phase forbidden
emission--line spectrum, and suggested tentative identifications of
the emission peaks primarily as lines of [Fe~II], [Co~II], and
[Co~III].  On the other hand, our Figure~1 shows that the day~25 and
day~56 spectra are similar in many respects, the main differences
being confined to the 6600 to 7200\ang\ region.  This suggests that an
intepretation of the day~56 spectrum in terms of permitted--line
P--Cygni profiles should be explored.

Figure~7 shows a comparison of the day~56 spectrum with a synthetic
spectrum that has \vphot\ = 2000 \kms, $T_{bb}=7000$~K, and the same five
ions as used for the day~12 and day~25 spectra. The ion--specific
input parameters are in Table~4.  In the synthetic spectrum the Fe~II,
Co~II, and Cr~II lines are forming between 2000 and 7000 \kms, Ca~II
between 2000 and 8000 \kms, and Na~I between 5000 and 8000 \kms, all
with flat optical--depth distributions.  (The fit was slightly
improved by using an excitation temperature of 5000~K for Fe~II,
Co~II, and Cr~II instead of the default value of 7000~K that was used
at previous epochs.)  In spite of its imperfections (especially in the
6600 to 7200\ang\ region) the fit in Figure~7 is good enough to
indicate that our present interpretation of the spectrum is correct.
At 56~days postmaximum SN~2002cx had not yet made the transition to
a nebular--phase spectrum.

\section{DISCUSSION}

Figures~3 and 4 strongly suggest the presence of weak Si~II and S~II
features in the near--maximum spectra of SN~2002cx, which strengthens
the connection between SN~2002cx and SN~1991T--like events.  The
resemblance of the sped up synthetic spectrum for SN~2002cx to the
observed spectrum of SN~1997br (Figure~4) raises several questions
including the following.  First, do the similar line optical depths in
the two events mean that the spectra were forming in similar
compositions at similar temperatures, although at different
velocities?  This question can be answered only by means of detailed
spectrum calculations (e.g., Lentz et~al. 2001; H\"oflich
et~al. 2002).  Second, the Si~II and S~II features are not detectable
in the day~16 spectrum of SN~2002cx, although they were conspicuous at
the same epoch (and even later epochs) of SNe~1991T and 1997br.  Did
SN~2002cx develop conspicuous Si~II and S~II features at some time
between day~$-1$ and day~12?  This of course cannot be answered until
a SN~2002cx--like event is observed at such phases.  Third, did the
near--maximum spectrum of SN~1997br contain all of the same weak
features as SN~2002cx, including the ones we have not identified, but
smeared out beyond recognition by the higher expansion velocities in
SN~1997br?  This question could be addressed by speeding up a {\bf
Synow} synthetic spectrum that accounts for all of the weak features
of SN~2002cx, if such a synthetic spectrum can be constructed.

For days~12, 25, and 56, the agreement between the synthetic and
observed spectra establishes that these are resonance--scattering
spectra dominated by lines of Fe~II and Co~II.  The relative optical
depths of the Fe~II and Co~II reference lines (2, 1.67, and 2 at the
three epochs) are constant within our fitting freedom.  The
line--formation velocities are low: 7000 to 9000 \kms, 5000 to 7000
\kms, and 2000 to 7000 \kms\ at the three epochs.  The outer boundary
of the line--formation region of 7000 \kms\ at days~25 and 56 can be
compared to values of 13,500, 12,500, and 10,000 \kms\ at which we
placed Fe~II optical--depth discontinuities in SNe~2000cx at 32 days
after maximum (Branch et~al. 2004), SN~1997br at 38 days after maximum
(Hatano et~al. 2002), and SN~1991T at 59 days postmaximum (Fisher et
al. 1999).  We have begun a comparative study of SN~Ia postmaximum
spectra to explore the extent to which the differences between
SN~2002cx and other SNe~Ia are due just to the low velocity of the
line-formation region, or also to different line optical depths.
Thanks to the low velocity of the line--formation region the SN~2002cx
postmaximum spectral features can be identified with more confidence
than in other SNe~Ia.  The comparative study will lead to improved
line identifications in other SNe~Ia, and clarify our understanding of
when the transition from the permitted--line to the forbidden--line
phase occurs.

Near maximum light SN~2002cx had normal colors and a high--excitation
SN~1991T--like spectrum, but a low expansion velocity and a low
luminosity.  After maximum the spectra rapidly developed strong
permitted lines of Fe~II and Co~II that persisted to at least 56 days
postmaximum.  At all epochs the velocities were low, which implies low
kinetic energy.  L03 discussed the properties of one dimensional (1D)
explosion models and concluded that no published model can account for
SN~2002cx.  One basic problem is that if practically all of the
ejected mass was burned from carbon and oxygen to iron--group
elements, which at first glance the spectra seem to suggest, then the
kinetic energy per unit mass and the expansion velocities of SN~2002cx
should be high rather than low.

Recently 3D explosion models, both deflagrations and delayed
detonations, have begun to appear.  According to Gamezo, Khokhlov, \&
Oran (2004), 3D deflagration models cannot account for normal SNe~Ia
because they produce too little kinetic energy, too little radioactive
nickel, and too much unburned carbon and oxygen at low velocity;
Gamezo et~al. argue that normal SNe~Ia require a delayed detonation.
Also, the composition structure of 3D deflagrations is too clumpy to
be able to produce deep Si~II absorptions (Thomas et~al. 2002), such
as are observed not only in normal SNe~Ia but also in the peculiar
SN~1991bg--like events.  However, SN~2002cx is quite unlike both
normal SNe~Ia and SN~1991bg--like events, and its unusual properties
may be consistent with a 3D deflagration model.  Its kinetic energy
and luminosity were low, and deep Si~II lines were not observed
(although we must keep in mind that the spectrum was not observed
between day~$-1$ and day~12).  And perhaps it is not true that
practically all of the carbon and oxygen was burnt to iron--peak
elements.  Unburned carbon and oxygen at low velocity, coexisting with
or beneath a substantial mass of iron--peak elements, may be difficult
to detect owing to the ``iron curtain'': ubiquitous line blocking by
Fe~II lines forming at similar or higher velocities (see Baron, Lentz,
\& Hauschildt 2003).  In a forthcoming paper we will return to the
issue of low--velocity carbon and oxygen in an analysis of
second--season spectra of SN~2002cx, obtained eight and nine months
after explosion.

Another possibility is that a substantial asymmetry is involved.
Kasen (2004) and Kasen et~al. (2004) have suggested that a normal
SN~Ia viewed down the hole in the ejecta caused by the presence of a
nondegenerate companion star (Marietta et~al. 2000) would have some
characteristics of SN~1991T--like events.  They mention that a
subluminous SN~1991bg--like event viewed down the hole might have some
of the properties of SN~2002cx.  This idea deserves further attention.

SN~2002cx was an observationally rare event (but perhaps not unique;
see Filippenko \& Chornock 2002 on SN~2003gq).  Observations of many
more SNe~Ia should reveal whether some events have properties between
those of SN~2002cx and other SNe~Ia, or that SN~2002cx--like events
comprise a distinct sub--type of SNe~Ia.

We are grateful to S.~Jha for helpful discussions.  This work has been
supported by NSF grants AST-0204771, AST-0307323, AST-0307894, and
NASA grant NAG 5-12127.

\clearpage     

\begin{figure}
\includegraphics[width=.7\textwidth,angle=270]{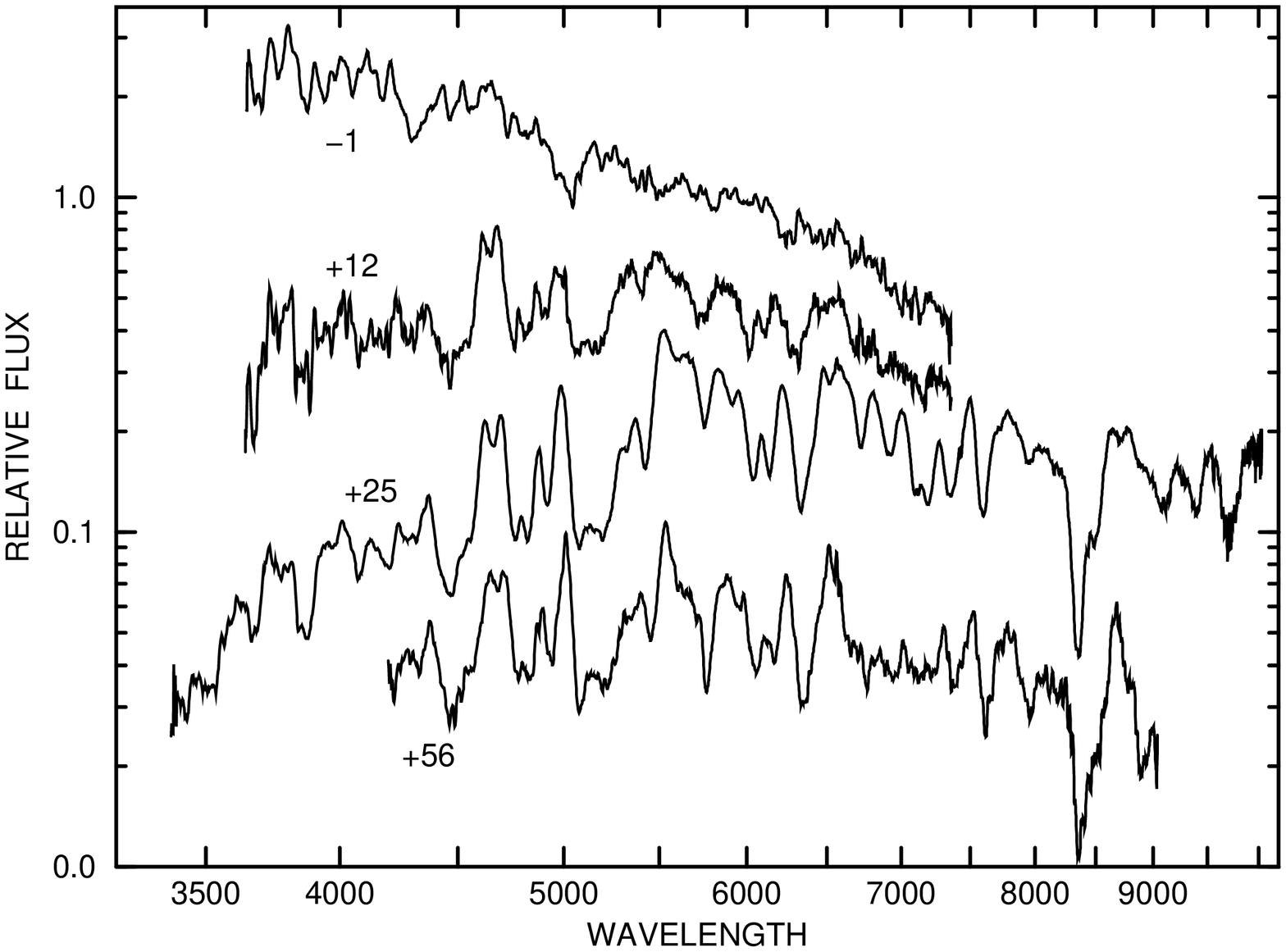}
%\plotone{ x02cx4sp.comploglog.eps}
\caption{Four spectra of SN~2002cx, from L03, are compared.  Epochs
  are in days with respect to the date of maximum brightness in the
  $B$ band, 2002 May~21.  The flux is per unit wavelength interval and
  the vertical displacement is arbitrary.  Spectra have been corrected
  for the redshift of the parent galaxy, $cz = 7184$~\kms.  No
  correction for interstellar reddening has been applied.}
\end{figure}

\begin{figure}
\includegraphics[width=.8\textwidth,angle=270]{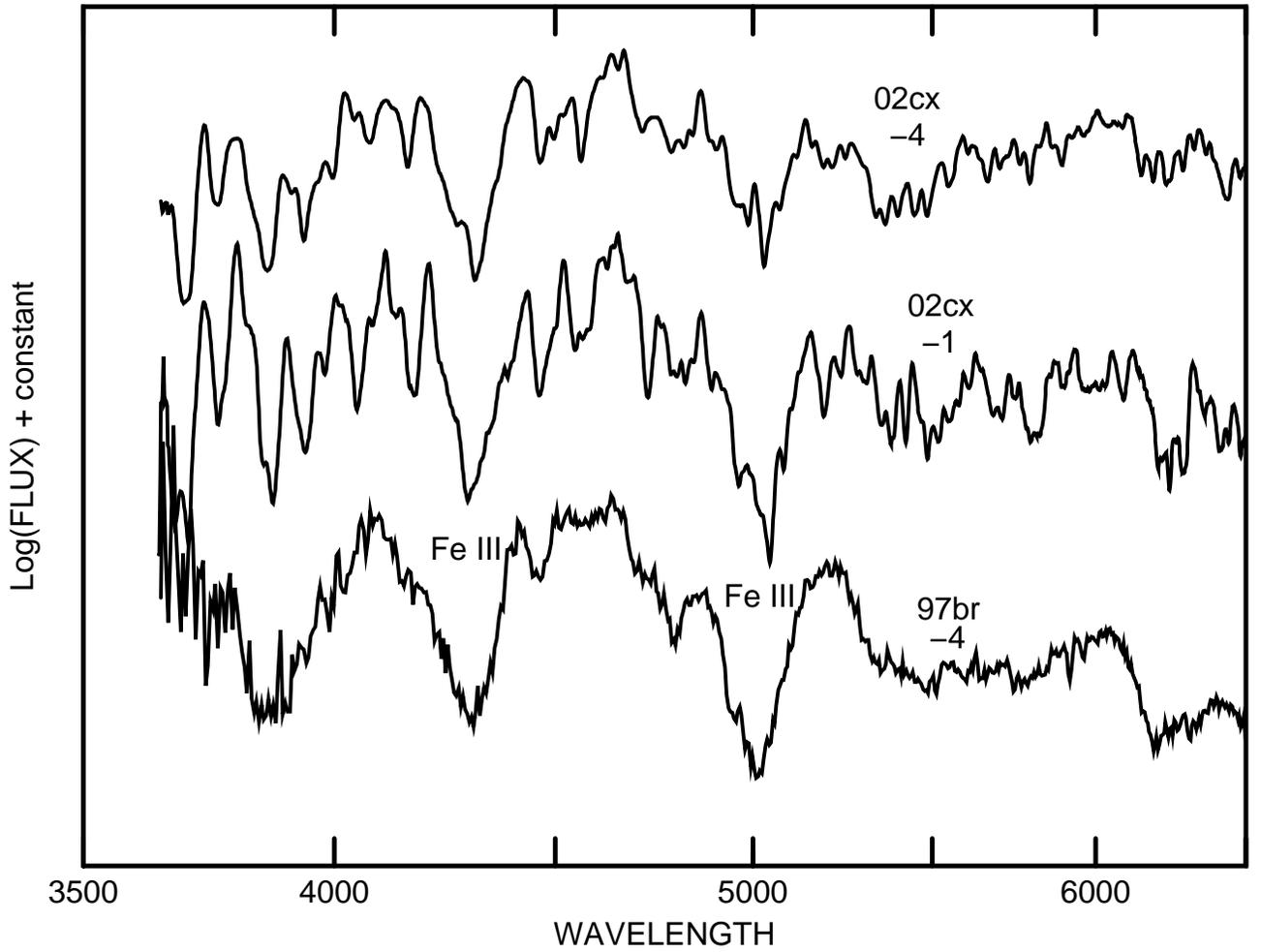}
\caption{The day~$-4$ and day~$-1$ spectra of SN~2002cx, and the
  day~$-4$ spectrum of SN~1997br, are compared.  The SN~1997br
  spectrum has been redshifted by 3000 \kms\ to align the Fe~III
  absorptions.  The flux is per unit frequency interval (so that the
  spectra are roughly flat rather than steeply sloped) and the
  vertical displacement is arbitrary.}
\end{figure}

\begin{figure}
\includegraphics[width=.8\textwidth,angle=270]{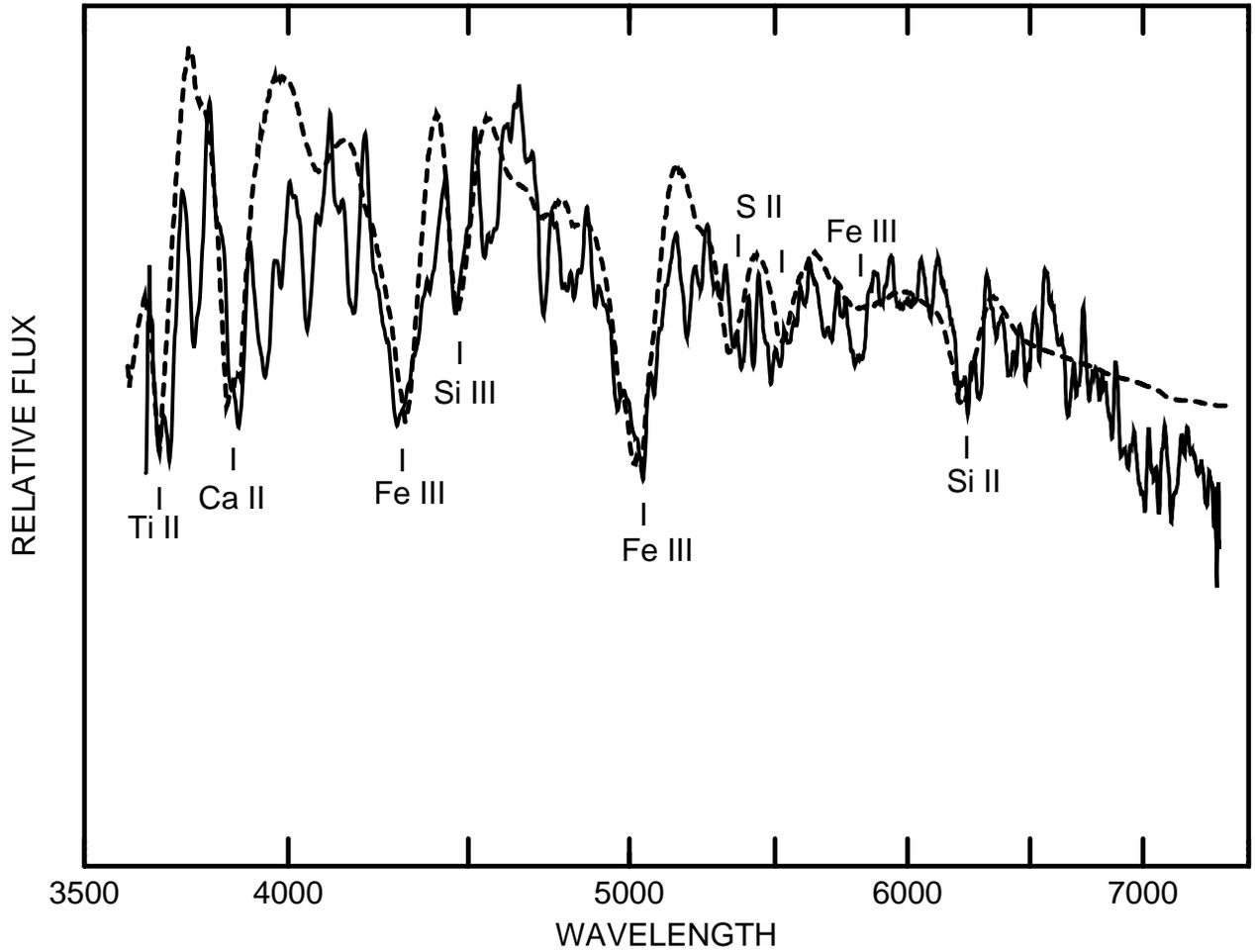}
\caption{The day~$-1$ spectrum of SN~2002cx ({\sl solid line}) is
  compared with a synthetic spectrum ({\sl dashed line}) that has
  $v_{phot}=7000$ \kms, $T_{bb}=14,000$~K, and contains lines of six
  ions. The flux is per unit frequency interval.  In this and
  subsequent figures the flux is linear rather than logarithmic.}
\end{figure}

\begin{figure}
\includegraphics[width=.8\textwidth,angle=270]{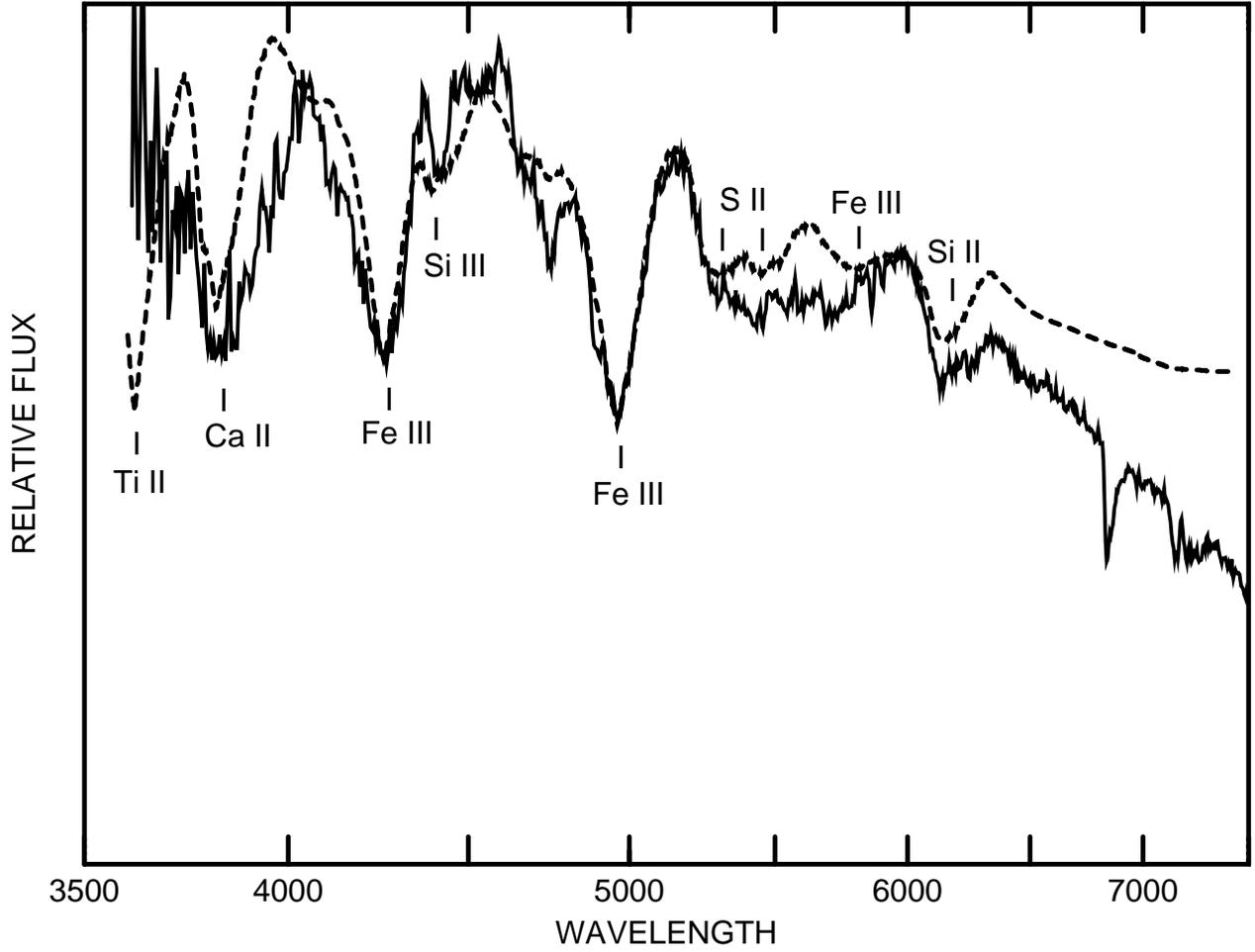}
\caption{The day~$-4$ spectrum of SN~1997br ({\sl solid line}) is
  compared with a synthetic spectrum ({\sl dashed line}) like that of
  Figure~3, for SN~2002cx, except that \vphot\ has been increased from
  7000 to 12,000 \kms.}
\end{figure}

\begin{figure}
\includegraphics[width=.8\textwidth,angle=270]{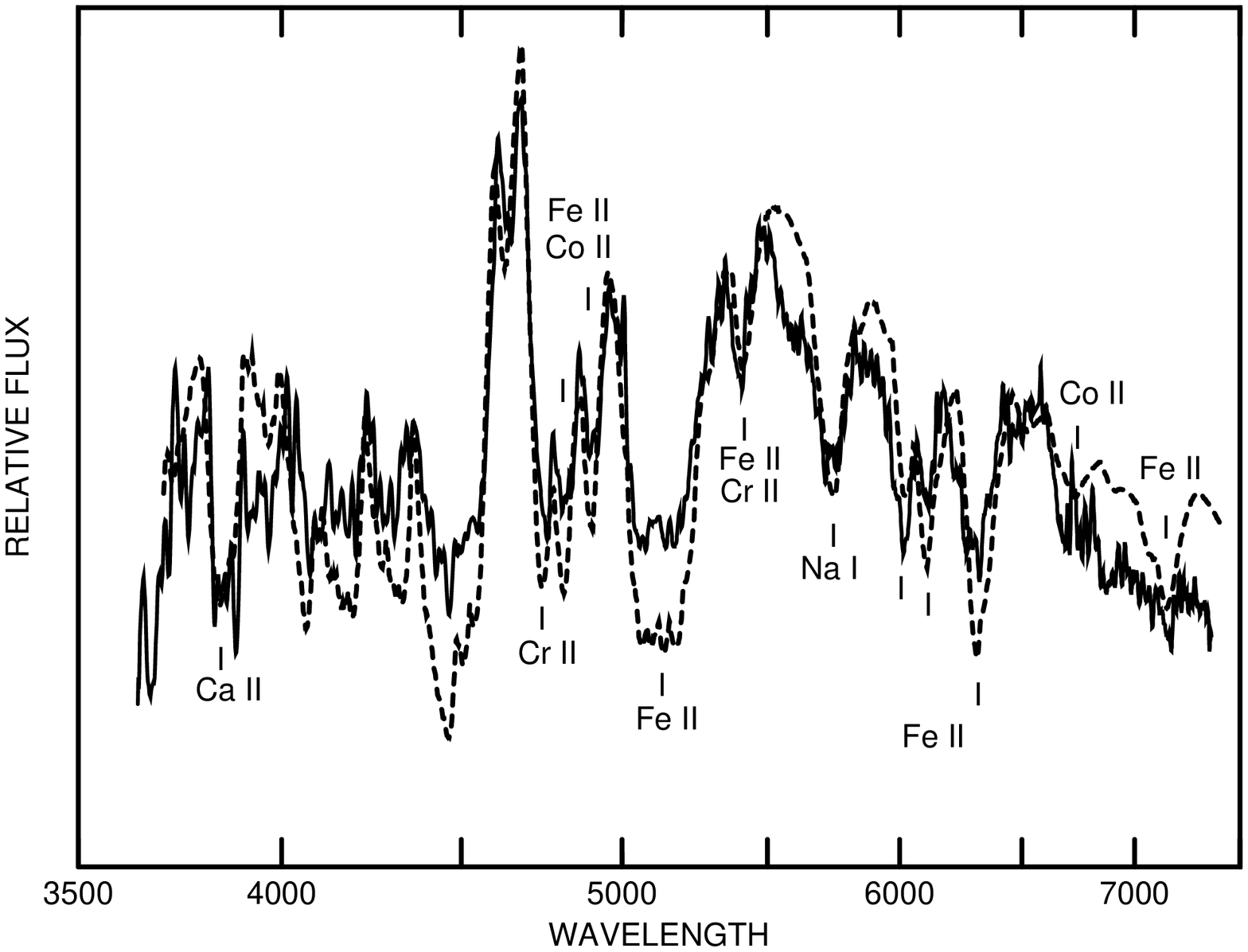}
\caption{The day~12 spectrum of SN~2002cx ({\sl solid line}) is
  compared with a synthetic spectrum ({\sl dashed line}) that has
  $v_{phot}=7000$ \kms, $T_{bb}=9000$~K, and contains lines of five
  ions.  The flux is per unit wavelength interval.}
\end{figure}

\begin{figure}
\includegraphics[width=.8\textwidth,angle=270]{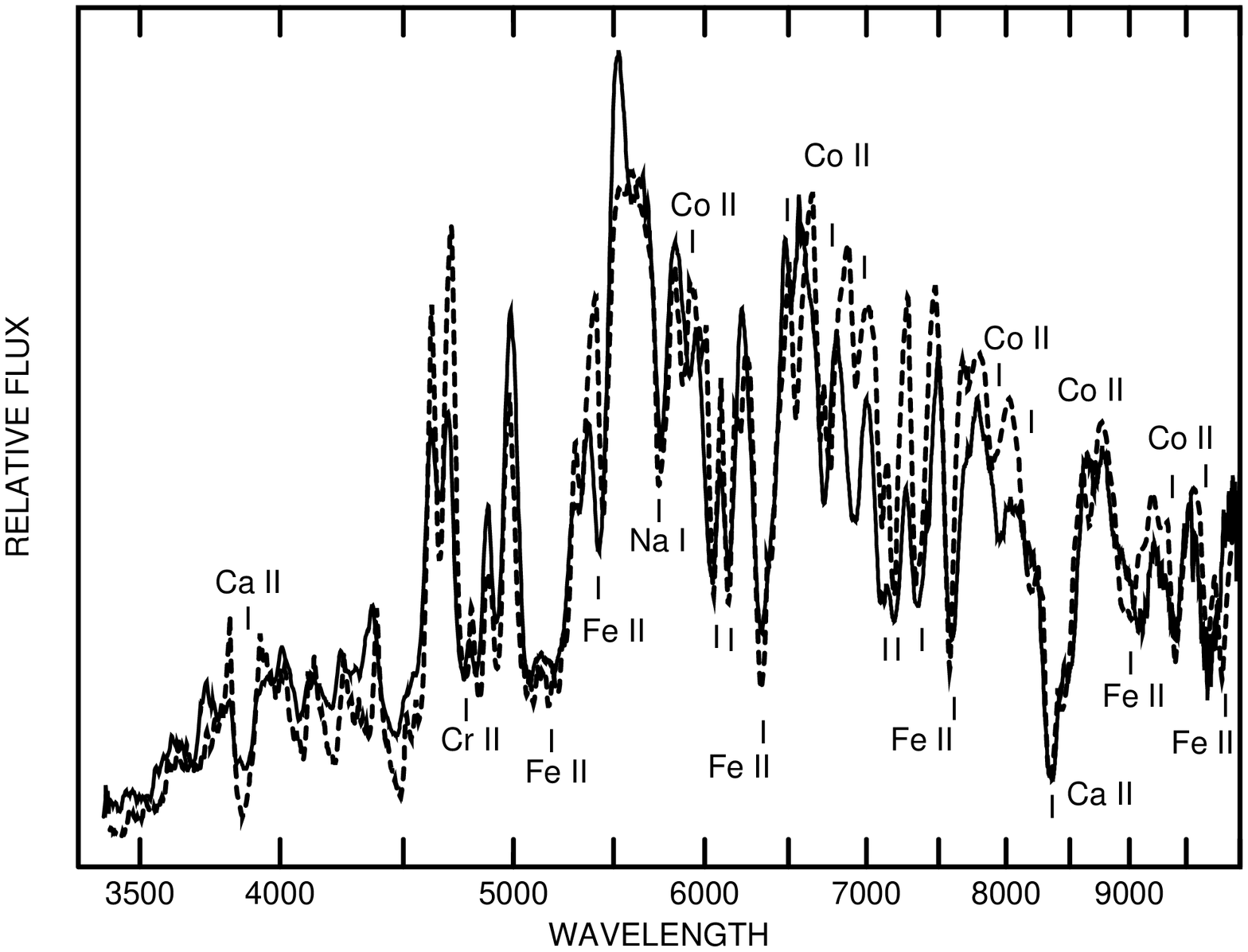}
\caption{The day~25 spectrum of SN~2002cx ({\sl solid line}) is
  compared with a synthetic spectrum ({\sl dashed line}) that has
  $v_{phot}=5000$ \kms, $T_{bb}=5500$~K, and contains lines of five
  ions.  The flux is per unit wavelength interval.}
\end{figure}

\begin{figure}
\includegraphics[width=.8\textwidth,angle=270]{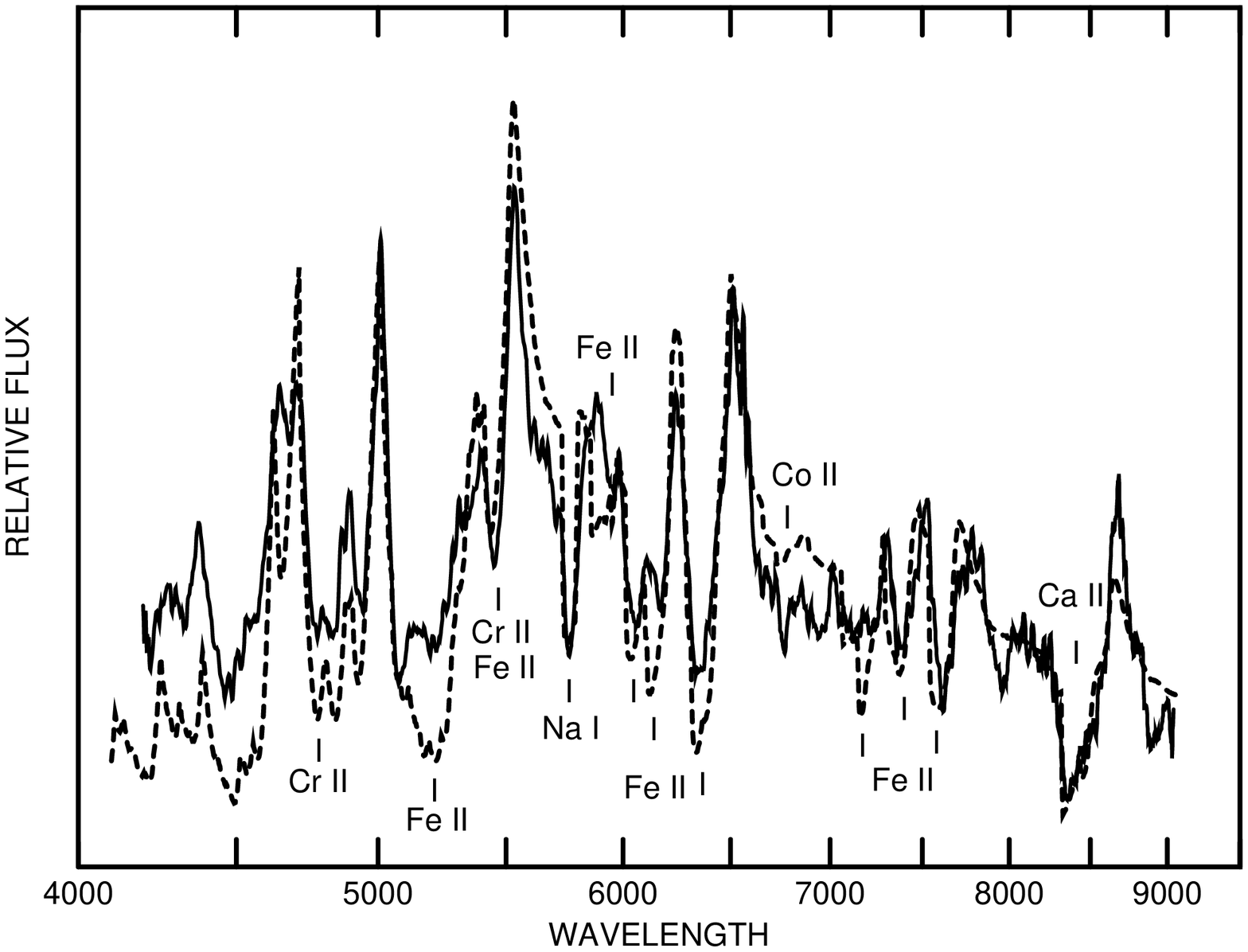}
\caption{The day~56 spectrum of SN~2002cx ({\sl solid line}) is
  compared with a synthetic spectrum ({\sl dashed line}) that has
  $v_{phot}=2000$ \kms, $T_{bb}=7000$~K, and contains lines of five
  ions.   The flux is per unit wavelength interval.}
\end{figure}

\clearpage
 
\begin{deluxetable}{lcccccr}
\footnotesize
\tablecaption{Input Parameters for Figure 3 (Day $-1$) \label{table1}}
\tablewidth{0pt}
\tablehead{
\colhead{ion} &
\colhead{$\lambda$(ref)} &
\colhead{$\tau$(ref)} & 
\colhead{$v_{min}$} &
\colhead{$v_{max}$} &
\colhead{$v_e$} &
\colhead{$T_{exc}$} \\
\colhead{} &
\colhead{(\AA)} &
\colhead{} &
\colhead{(\kms)} &
\colhead{(\kms)} &
\colhead{(\kms)} &
\colhead{(K)} 
} 
\startdata
Si II  &$\lambda$6347 &0.5  & 7000  &$\infty$ &1000    & 7000 \\
Si III &$\lambda$4553 &0.7  & 7000  &$\infty$ &1000    &13,000 \\
S II   &$\lambda$5454 &0.5  & 7000  &$\infty$ &1000    & 9000 \\
Ca II  &$\lambda$3934 &3.0  & 7000  &$\infty$ &1000    & 5000 \\
Ti II  &$\lambda$4550 &0.1  & 7000  &$\infty$ &1000    & 5000 \\
Fe III &$\lambda$4420 &1.5  & 7000  &$\infty$ &1000    &13,000 \\
\enddata
\end{deluxetable}
\clearpage

\begin{deluxetable}{lcccccr}
\footnotesize
\tablecaption{Input Parameters for Figure 5 (Day 12) \label{table2}}
\tablewidth{0pt}
\tablehead{
\colhead{ion} &
\colhead{$\lambda$(ref)} &
\colhead{$\tau$(ref)} & 
\colhead{$v_{min}$} &
\colhead{$v_{max}$} &
\colhead{$v_e$} &
\colhead{$T_{exc}$} \\
\colhead{} &
\colhead{(\AA)} &
\colhead{} &
\colhead{(\kms)} &
\colhead{(\kms)} &
\colhead{(\kms)} &
\colhead{(K)}
} 
\startdata
Na I   &$\lambda$5890 &0.5   &  7000 &  9000    &$\infty$   & 5000 \\
Na I   &$\lambda$5890 &0.5   &  9000 &13,000    &  2000    & 5000 \\
Ca II  &$\lambda$3934 &2.0   &  7000 &  9000    &$\infty$    & 5000 \\
Ca II  &$\lambda$3934 &2.0   &  9000 &13,000    &  2000    & 5000 \\
Cr II  &$\lambda$4240 &3.0   &  7000 &  9000    &$\infty$    & 7000 \\
Fe II  &$\lambda$5018 &40    &  7000 &  9000    &$\infty$    & 7000 \\
Co II  &$\lambda$4161 &20    &  7000 &  9000    &$\infty$    & 7000 \\
\enddata
\end{deluxetable}
\clearpage

\begin{deluxetable}{lcccccr}
\footnotesize
\tablecaption{Input Parameters for Figure 6 (Day 25) \label{table3}}
\tablewidth{0pt}
\tablehead{
\colhead{ion} &
\colhead{$\lambda$(ref)} &
\colhead{$\tau$(ref)} & 
\colhead{$v_{min}$} &
\colhead{$v_{max}$} &
\colhead{$v_e$} &
\colhead{$T_{exc}$} \\
\colhead{} &
\colhead{(\AA)} &
\colhead{} &
\colhead{(\kms)} &
\colhead{(\kms)} &
\colhead{(\kms)} &
\colhead{(K)}
} 
\startdata
Na I   &$\lambda$5890 &0.4   &  5000 &  9000    &$\infty$    & 5000 \\
Ca II  &$\lambda$3934 &300   &  5000 &  9000    &$\infty$  & 5000 \\
Cr II  &$\lambda$4240 &3.0   &  5000 &  7000    &$\infty$    & 7000 \\
Fe II  &$\lambda$5018 &100   &  5000 &  7000    &$\infty$    & 7000 \\
Co II  &$\lambda$4161 &60    &  5000 &  7000    &$\infty$    & 7000 \\
\enddata
\end{deluxetable}
\clearpage

\begin{deluxetable}{lcccccr}
\footnotesize
\tablecaption{Input Parameters for Figure 7 (Day 56) \label{table4}}
\tablewidth{0pt}
\tablehead{
\colhead{ion} &
\colhead{$\lambda$(ref)} &
\colhead{$\tau$(ref)} & 
\colhead{$v_{min}$} &
\colhead{$v_{max}$} &
\colhead{$v_e$} &
\colhead{$T_{exc}$} \\
\colhead{} &
\colhead{(\AA)} &
\colhead{} &
\colhead{(\kms)} &
\colhead{(\kms)} &
\colhead{(\kms)} &
\colhead{(K)}
} 
\startdata
Na I   &$\lambda$5890 &0.5   &  5000 &  8000    &$\infty$    & 5000 \\
Ca II  &$\lambda$3934 &120   &  2000 &  8000    &$\infty$    & 5000 \\
Cr II  &$\lambda$4240 &9.0   &  2000 &  7000    &$\infty$    & 5000 \\
Fe II  &$\lambda$5018 &80    &  2000 &  7000    &$\infty$    & 5000 \\
Co II  &$\lambda$4161 &40    &  2000 &  7000    &$\infty$    & 5000 \\
\enddata
\end{deluxetable}
\clearpage

\end{document}